\documentclass[9pt,twocolumn,twoside]{opticajnl}
\journal{ol} % Choose journal (ao,jocn,josaa,josab,ol,optica,pr)

\setboolean{shortarticle}{true}
\usepackage[labelformat=simple]{subcaption}

\DeclareCaptionLabelFormat{subcaptionlabel}{\normalfont(\textbf{#2}\normalfont)}
\usepackage{cleveref}
\crefname{figure}{Fig.}{Figs.}
\Crefname{figure}{Figure}{Figures}
\crefname{equation}{Eq.}{Eqs.}
\Crefname{equation}{Equation}{Equations}
\crefrangelabelformat{equation}{(#3#1#4)-(#5#2#6)}

\usepackage{amsmath}
\usepackage{upgreek}

\usepackage{lineno}
\usepackage{titlesec}
\titlespacing*{\subsection}{0pt}{2pt}{0pt}
\title{Synchrotron-based X-ray Fluorescence Ghost Imaging}

\author[1,2]{Mathieu Manni}
\author[2]{Adi Ben-Yehuda}
\author[2]{Yishay Klein}
\author[1]{Bratislav Lukic}
\author[3]{Andrew Kingston}
\author[1]{Alexander Rack}
\author[2,*]{Sharon Shwartz}
\author[1,4,**]{Nicola Vigan\`{o}}

\affil[1]{ESRF --- The European Synchrotron, Grenoble, 38043, France}
\affil[2]{Physics Department and Institute of Nanotechnology and Advanced Materials, Bar Ilan University, Ramat Gan, 52900, Israel}
\affil[3]{Department of Materials Physics, Research School of Physics, The Australian National University, Canberra, ACT 2601, Australia}
\affil[4]{IRIG-MEM, CEA, Universit\'{e} Grenoble Alpes, Grenoble, 38000, France}

\affil[*]{Co-corresponding author: sharon.shwartz@biu.ac.il}
\affil[**]{Co-corresponding author: nicola.vigano@cea.fr}

\begin{abstract}
X-ray Fluorescence Ghost Imaging (XRF-GI) was recently demonstrated for x-ray lab sources. It has the potential to reduce acquisition time and deposited dose by choosing their trade-off with spatial resolution, while alleviating the focusing constraints of the probing beam. Here, we demonstrate the realization of synchrotron-based XRF-GI: We present both an adapted experimental setup and its corresponding required computational technique to process the data. This extends the above-mentioned potential advantages of GI to synchrotron XRF imaging. In addition, it enables new strategies to improve resilience against drifts at all scales, and the study of previously inaccessible samples, such as liquids.
\end{abstract}

\setboolean{displaycopyright}{true}

\begin{document}

\maketitle

\subsection{Introduction}
X-ray fluorescence (XRF) is used in a wide variety of experimental techniques, including two-dimensional and three-dimensional chemical mapping spanning a large range of scales. It is used for sample characterization in a multitude of application areas, including material science~\cite{cit-material-science}, chemistry~\cite{cit-batteries}, and cultural heritage~\cite{cit-cultural-heritage}.  % , archaeology~\cite{cit-archaeology}
The chemical sensitivity of XRF is achieved through the excitation of core electrons: When the excited atoms return to lower excitation states, they emit secondary photons with characteristic energies, that uniquely depend on the atomic number. Energy-resolving detectors (routinely used in x-ray measurements), can discriminate these photons with sufficient energy resolution, to identify the element from which they originated.
XRF imaging is usually achieved by scanning the samples with a focused beam (pencil beam, PB), and by collecting the emitted XRF signal (spectrum) with single-pixel energy-resolving detectors. The collected XRF spectra are then processed, to fit the measured local chemical composition~\cite{Sole2007}.
\\
% Introduce problems and pitfalls of XRF imaging 
PB acquisitions require raster-scanning (sequentially) of all the points in the field of view (FoV), by transversely displacing the sample. Each exposure has an associated XRF spectrum, which corresponds to one pixel in the resulting spectral image.
Synchrotron radiation, compared to x-ray tube-based sources, is characterized by higher photon flux and (on long beamlines) spatial coherence. This enables reaching x-ray beam waists of tens of nm with a high photon flux~\cite{daSilva2017}, which, in turn, enables the study of micro- and nano-structured samples with unrivaled speeds, compared to laboratory sources.
As it delivers a high radiation dose rate per unit-surface per unit-time, it can cause serious localized damage and deformation in sensitive samples. Heat can also contribute to positional drifts and uncertainties, potentially leading to image degradation.
%GI can alleviate these effects: The dose per unit-surface per unit-time is reduced by the ratio of the focused beam waist divided by the FoV.
\\
In contrast to PB, classical Ghost Imaging (GI) acquisitions illuminate the entire FoV. Differently from other full-field techniques~\cite{Vasin2007, Soltau2023}, it does so with different structured beams at each exposure~\cite{Moreau2018}. For x-rays, these structured beams are usually obtained by inserting non-configurable transversely-displaced structuring masks in the beam, to encode the spatial information in the acquired GI signals. In the specific case of XRF-GI, the XRF detector records the spectrum associated with each illumination pattern. The XRF energy emission lines corresponding to different chemical elements are reconstructed into spatially resolved maps, using computational imaging algorithms.
GI acquires spatial information on the whole FoV at each realization. Thanks to the inherent compressibility of natural images, it is possible to acquire fewer realizations than the number of pixels in the reconstructed image, leading to reduced dose deposition, which is not possible with PB scans~\cite{Lane2020}.
The sole translation of the masks in XRF-GI, compared to the translation of the samples for PB, enables the study of previously inaccessible samples like liquids, that cannot move during measurements.
Moreover, by spreading the beam flux over the entire FoV (as opposed to just the focal spot in PB), XRF-GI could offer more efficient mitigation of dose effects (e.g. easier cooling) and reduce the localized radiation-induced damage.
Despite having first developed XRF-GI on laboratory equipment~\cite{Klein2022}, the transposition to synchrotron beamlines is highly desirable, even though it comes with its own challenges.
\begin{figure}[t]
    \centering
    \begin{subfigure}[b]{\linewidth}
        \setlength{\abovecaptionskip}{4pt}
        \setlength{\belowcaptionskip}{-2pt}
        \includegraphics[width=\linewidth]{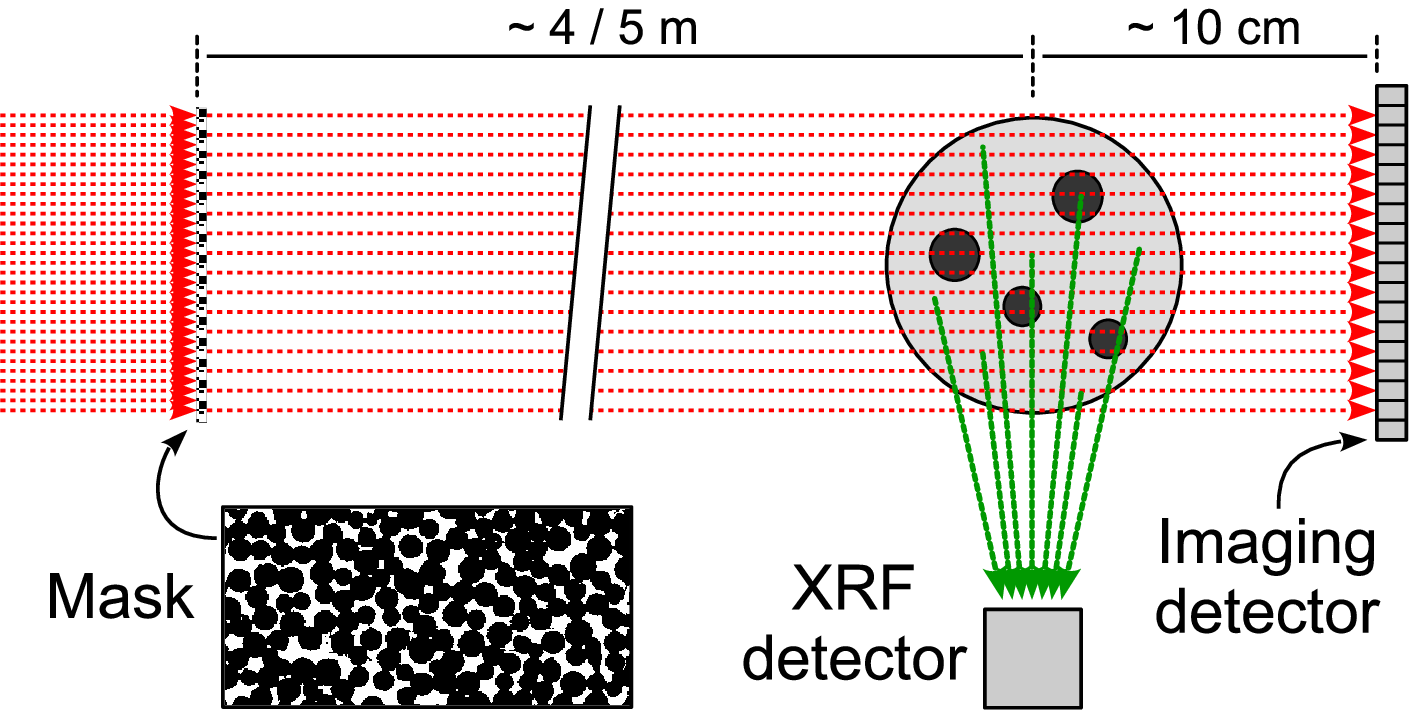}
        \subcaption{Schematic setup}
        \label{fig:geom:scheme}
    \end{subfigure}
    \\
    \vspace{0.2cm}
    \begin{subfigure}[b]{0.42\linewidth}
        \setlength{\abovecaptionskip}{4pt}
        \setlength{\belowcaptionskip}{-2pt}
        \includegraphics[width=\linewidth]{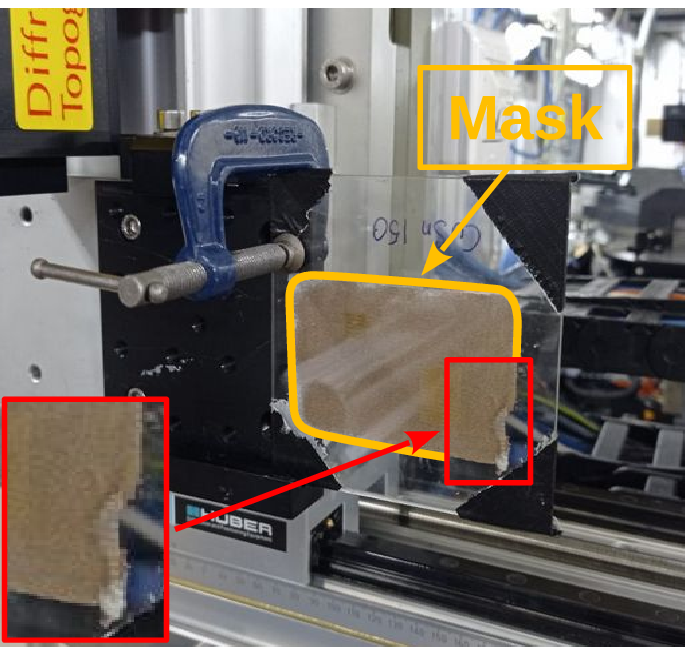}
        \subcaption{Mask}
        \label{fig:geom:mask}
    \end{subfigure}
    \hfill
    \begin{subfigure}[b]{0.56\linewidth}
        \setlength{\abovecaptionskip}{4pt}
        \setlength{\belowcaptionskip}{-2pt}
        \includegraphics[width=\linewidth]{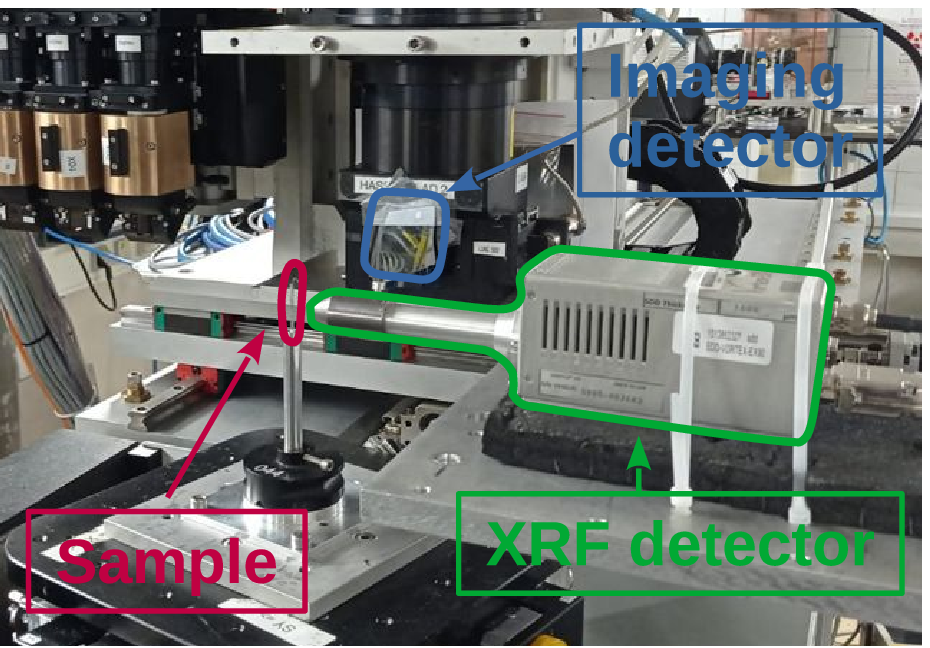}
        \subcaption{Sample \& detectors}
        \label{fig:geom:sample}
    \end{subfigure}

    \caption{(\textbf{\subref{fig:geom:scheme}})~Schematic synchrotron XRF-GI setup; (\textbf{\subref{fig:geom:mask}})~CuSn mask, whose corner is shown in the inset on the bottom left; and, (\textbf{\subref{fig:geom:sample}})~sample, imaging detector with Si attenuation wafers, and XRF detector.}
    \label{fig:geometry}
\end{figure}
\\
% The lab-based XRF-GI demonstrated in~\cite{Klein2022} assumes a constant flux from the x-ray source (or the flux to be directly measurable upstream to the sample), and the positions of the masks to not undergo drifts during the measurement of the XRF signal. These are valid assumptions for lab x-ray sources coupled with detectors with pixel sizes in the order of the tens of $\upmu m$.
X-ray GI image reconstructions require the mask shape and positions, and the incident beam intensity to be known.
% If the beam intensity is not correctly measured, its changes are especially detrimental when they are higher than the intensity variation introduced by the masks.
In contrast to x-ray tubes (used in laboratory x-ray setups) that are characterized by rather stable emission fluxes over time, synchrotron sources exhibit beam intensity decay. Even in top-up mode, flux variations of up to 10\% can be observed. Beam monitors are usually placed upstream of the masks. However, the masks create unknown and variable attenuation between the diode and the sample, which renders calibration of the XRF signal difficult or impossible.
The mask shapes can be known by: Measuring them before the scans or knowing their design from fabrication (Computational GI); splitting the structured beam with a beam-splitter positioned after the masks, and measuring the beam that does not impinge on the sample (Classical GI)~\cite{Bennink2002, Pelliccia2016, Gatti2004}. On synchrotron beamlines, cameras with sub-$\upmu$m or few $\upmu$m pixel sizes are routinely used. With slight instability of the beam direction and sample drifts, misalignment artifacts are more likely to manifest, which negatively impacts both the above-mentioned approaches. %  (e.g. drifts of reflecting x-ray optics visible due to the long lever arm at a beamline)
\\
% Describe what this paper proposes and its impact
Here, we demonstrate a synchrotron-based XRF-GI implementation that does not require knowing or measuring the masks in advance, to split the structured beam, calibrating with high-precision the positions of the masks, nor using a beam monitor to track possible incident beam intensity variations.

\subsection{Experimental setup}
The experimental setup is presented in~\cref{fig:geometry}.
A large structuring mask is positioned on a translation stage in the x-ray beam upstream of the sample. At each GI realization, the synchrotron x-ray beam only illuminates a portion of the mask, producing the corresponding illuminating beam shape. Different beam shapes are obtained by translating the mask transversely with respect to the incoming beam.
A 2D imaging detector is positioned downstream of the sample, and an XRF single-pixel detector is positioned next to the sample, perpendicular to the incoming beam direction.
We demonstrate our setup on the ID19 beamline of the ESRF --- The European Synchrotron, using a $1 \times 2 ~ $mm$^2$ incident beam size. The imaging detector is a so-called Hasselblad system, with two identical lenses (100 mm focal length) in tandem configuration (giving $\sim \times 1$ magnification), with a $500 \, \upmu$m LuAG:Ce scintillator and a ``Dimax pco.edge 5.5'' camera. It is positioned $\sim$~10~cm downstream of the sample, with an effective pixel size of $6 \, \upmu$m.
A large mask, composed of a mono-layer of randomly distributed CuSn (bronze) spheres (average diameter $\sim 50 \, \upmu$m and maximum diameter  $\sim 150 \, \upmu$m), is positioned $\sim 4 - 5$~m upstream of the sample. The incoming beam energy is $\sim 26$~keV, given by the beamline's single-harmonic undulator (type: u13), without the use of a monochromator. The XRF detector is a ``Hitachi Vortex 90EX'' single-pixel detector, controlled by a ``XIA FalconX'' module.
The sample is composed of two Cu flattened wires and one Fe $50 \, \upmu$m thick wire, stored in a capillary. The capillary is mostly made of plastic, with heavier trace elements.

\subsection{Acquisition procedure}
\begin{figure}[t]
    \centering
    \begin{subfigure}[b]{0.49\linewidth}
        \setlength{\abovecaptionskip}{4pt}
        \setlength{\belowcaptionskip}{-2pt}
        \includegraphics[width=\linewidth]{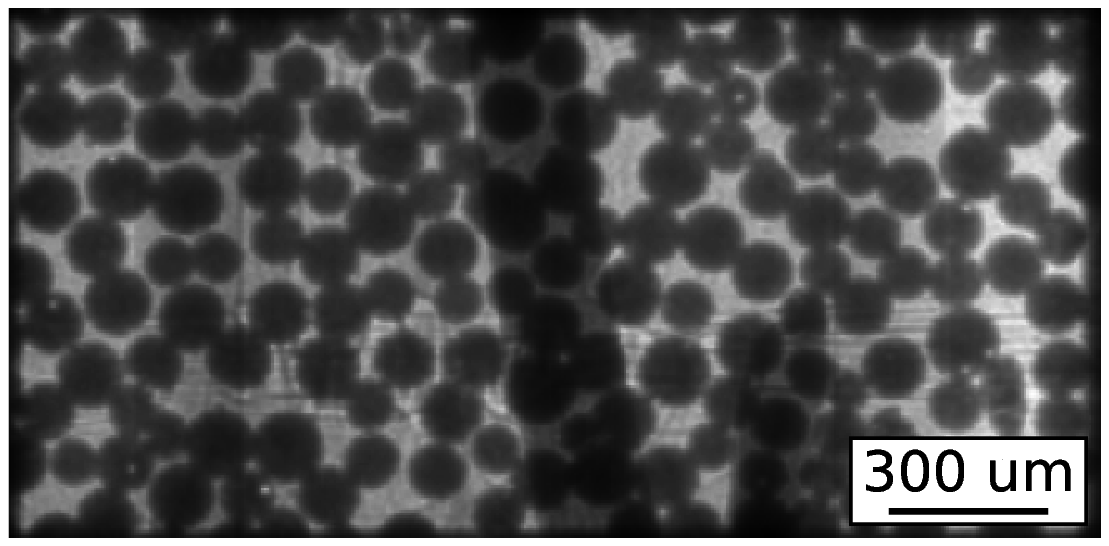}
        \subcaption{Transmission}
        \label{fig:eigff:transmission}
    \end{subfigure}
    \hfill
    \begin{subfigure}[b]{0.49\linewidth}
        \setlength{\abovecaptionskip}{4pt}
        \setlength{\belowcaptionskip}{-2pt}
        \includegraphics[width=\linewidth]{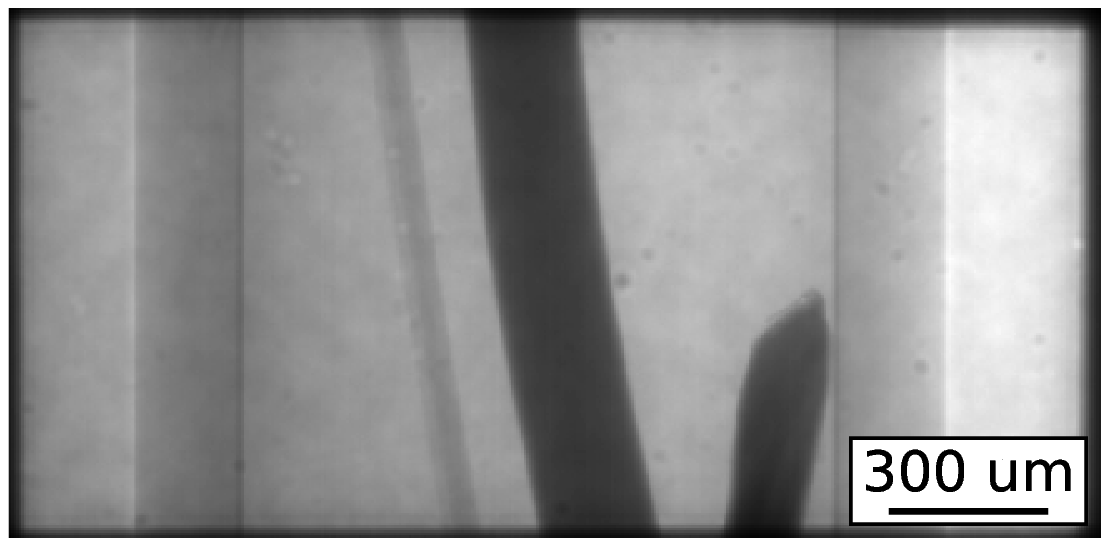}
        \subcaption{Eigen-flat-field}
        \label{fig:eigff:flatfield}
    \end{subfigure}
    \\
    \vspace{0.2cm}
    \begin{subfigure}[b]{0.49\linewidth}
        \setlength{\abovecaptionskip}{4pt}
        \setlength{\belowcaptionskip}{-2pt}
        \includegraphics[width=\linewidth]{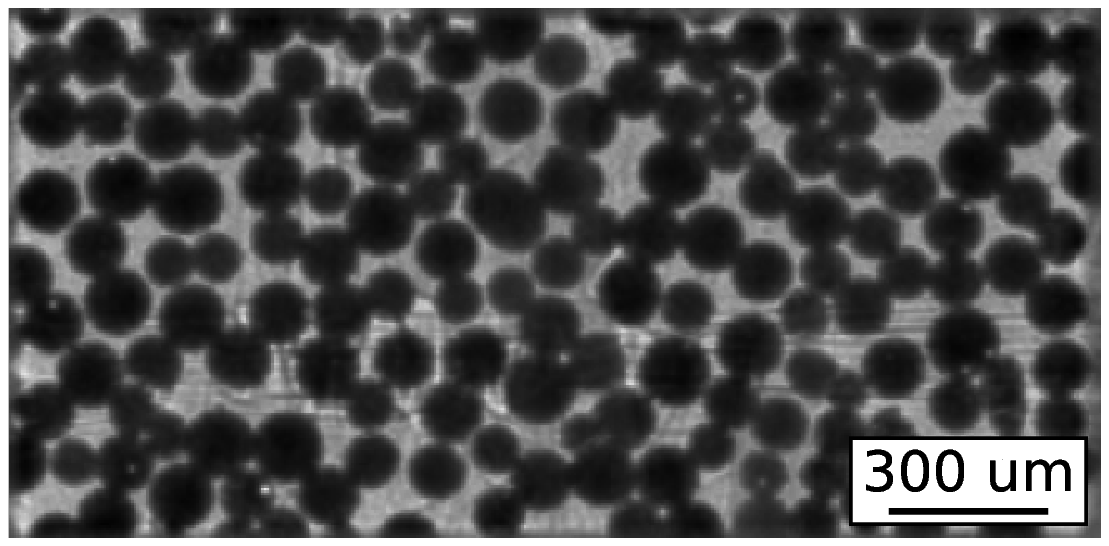}
        \subcaption{Structured beam}
        \label{fig:eigff:clean}
    \end{subfigure}
    \hfill
    \begin{subfigure}[b]{0.49\linewidth}
        \setlength{\abovecaptionskip}{4pt}
        \setlength{\belowcaptionskip}{-2pt}
        \includegraphics[width=\linewidth]{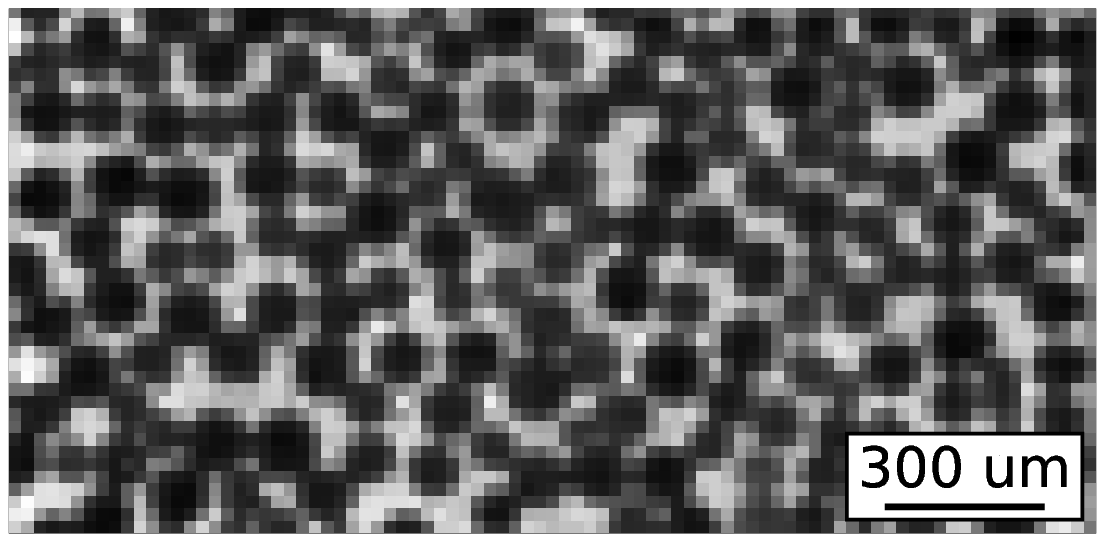}
        \subcaption{Binned structured beam}
        \label{fig:eigff:binned}
    \end{subfigure}

    \caption{(\textbf{\subref{fig:eigff:transmission}})~Transmission image of the sample with the mask; (\textbf{\subref{fig:eigff:flatfield}})~eigen-flat-field computed from the stack of transmission images; (\textbf{\subref{fig:eigff:clean}})~the extracted mask from transmission (normalized and mean-subtracted); and, (\textbf{\subref{fig:eigff:binned}})~its $4 \times 4$ binned version. The intensity gradient in~(\textbf{\subref{fig:eigff:transmission}}) and~(\textbf{\subref{fig:eigff:flatfield}}) is due to the fact that the sample is not correctly positioned in the center of the beam. This effect is correctly taken into account by the eigen-flat-field subtraction when retrieving~(\textbf{\subref{fig:eigff:clean}}).}
    \label{fig:eigff}
\end{figure}
In~\cite{Klein2022}, the imaging and XRF detectors acquired their respective signals separately. The sample needed to be removed from the FoV to acquire the beam structures. Thus, the same scan was performed two times, one with the sample in the FoV, using the XRF detector, and one without the sample in the FoV, using the imaging detector. This exposed one of the two measurements to positioning and flux estimation errors with respect to the other.
In this new implementation, the two detectors simultaneously acquire their respective signals (for the same time duration). In other words, each XRF signal is concurrently acquired with its corresponding beam structure, in the same flux and positioning conditions, and the imaging detector records the transmitted signal through both sample and mask for each GI realization.
We then separate the beam structure from the sample shape computationally, during the data processing.
Both implementations require measuring the intensity distribution of the impinging beam on the mask. This is done either at the beginning or the end of the scan.
\begin{figure*}[t]
    \centering
    \begin{subfigure}[b]{0.32\linewidth}
        \setlength{\abovecaptionskip}{4pt}
        \setlength{\belowcaptionskip}{-2pt}
        \includegraphics[width=\linewidth]{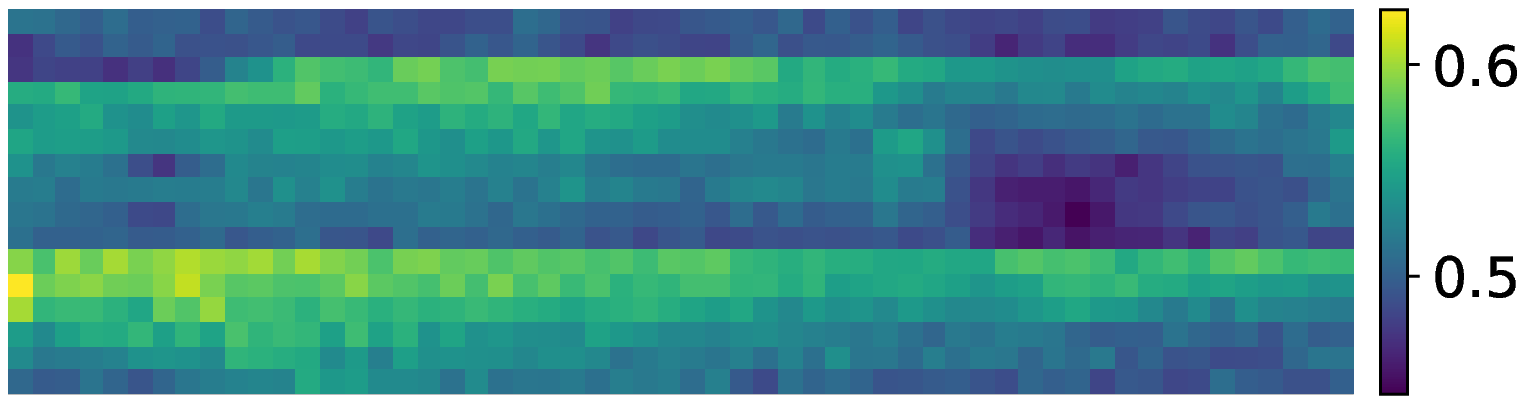}
        \subcaption{Eigen-flat-field intensity}
        \label{fig:data:norm}
    \end{subfigure}
    \hfill
    \begin{subfigure}[b]{0.32\linewidth}
        \setlength{\abovecaptionskip}{4pt}
        \setlength{\belowcaptionskip}{-2pt}
        \includegraphics[width=\linewidth]{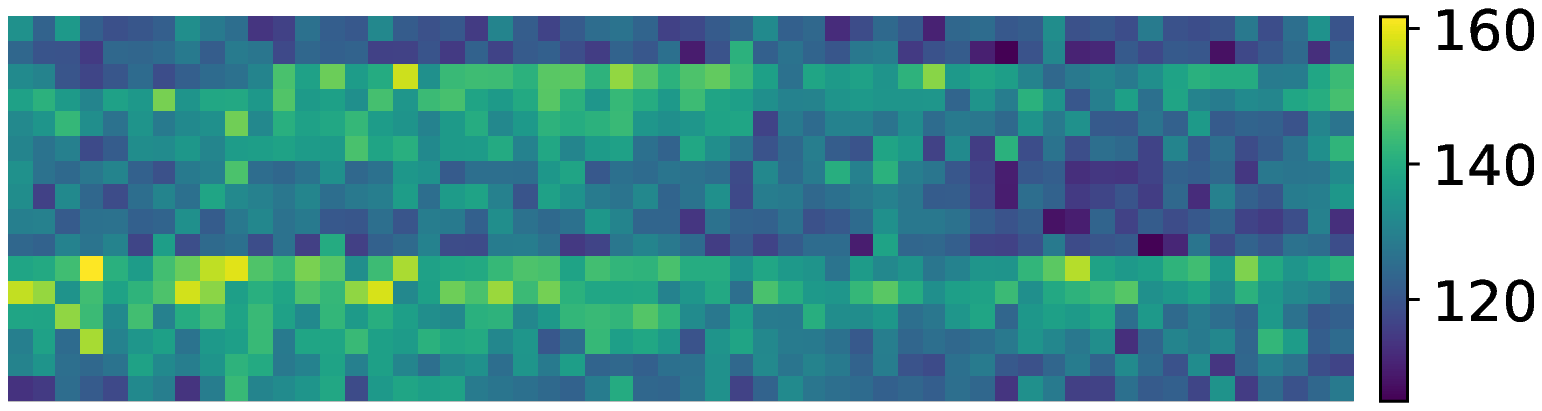}
        \subcaption{Acquired data intensity (Fe $K_\alpha$)}
        \label{fig:data:before}
    \end{subfigure}
    \hfill
    \begin{subfigure}[b]{0.32\linewidth}
        \setlength{\abovecaptionskip}{4pt}
        \setlength{\belowcaptionskip}{-2pt}
        \includegraphics[width=\linewidth]{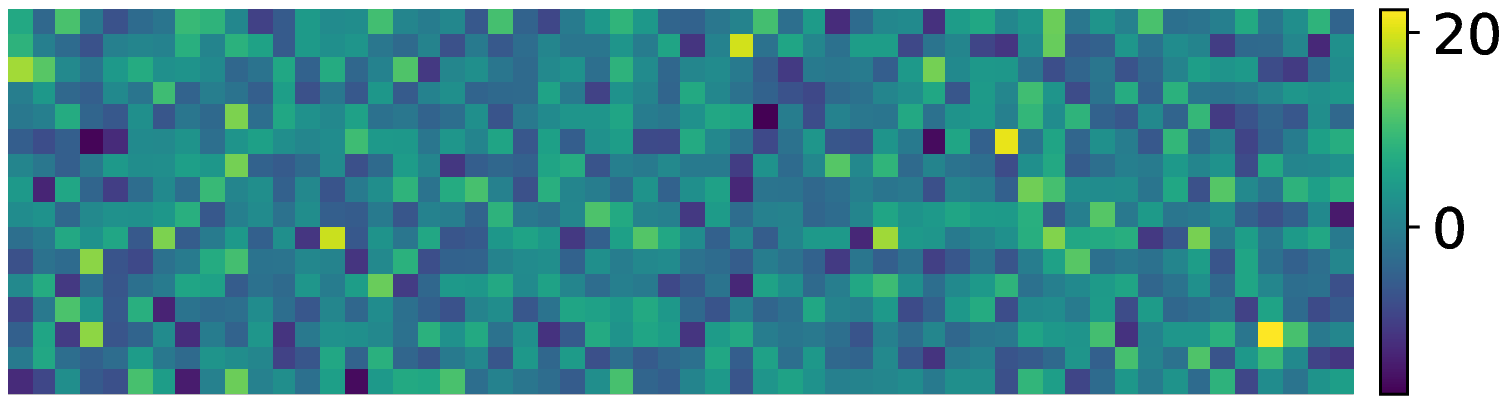}
        \subcaption{Normalized and mean-subtracted (Fe $K_\alpha$)}
        \label{fig:data:after}
    \end{subfigure}

    \caption{(\textbf{\subref{fig:data:norm}})~Integrated intensities of the eigen-flat-fields (\cref{fig:eigff:flatfield}); (\textbf{\subref{fig:data:before}})~photon counts for each GI realization of the Fe $K_\alpha$ line; and, (\textbf{\subref{fig:data:after}})~normalized and mean-subtracted intensities for the Fe $K_\alpha$ line (\textbf{\subref{fig:data:before}}) by the corresponding eigen-flat-field intensities (\textbf{\subref{fig:data:norm}}). Each point corresponds to one GI realization, for the 16 and 56 different positions vertically and horizontally respectively.}
    \label{fig:data}
\end{figure*}
\begin{figure*}[ht]
\centering
\includegraphics[width=\linewidth]{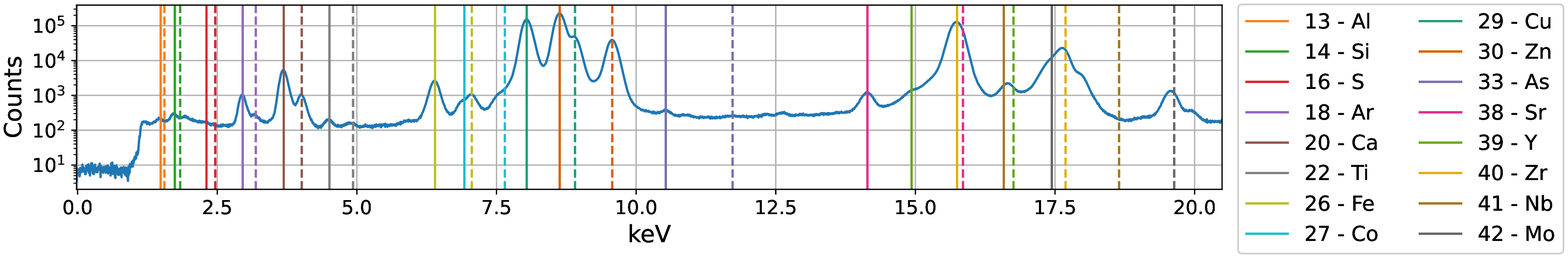}
\caption{XRF spectrum of the studied sample. Solid and dashed lines indicate $K_\alpha$ and $K_\beta$ emission lines, respectively.}
\label{fig:res:spectrum}
\end{figure*}
\\
The required exposure times of the high-resolution imaging and XRF detectors are usually different: The imaging detector is exposed to the direct (intense) beam, and therefore saturates well before the required SNR is met for the XRF signal. We circumvent this by acquiring many shorter acquisitions (e.g. exposures of 0.1~s each) for each same mask position and accumulating them.
Here, we translated the large mask both vertically and horizontally, acquiring 16 and 56 positions respectively, at a 5 mm translation step size. This amounts to 896 total GI realizations. We exposed 32 times for each GI realization for 0.1~s, amounting to 3.2~s of cumulative realization exposure time.
% By storing the individual sub-exposures for each mask position, it is also possible to evaluate the reconstruction quality degradation for shorter exposures, without the need to re-run the acquisition.

\subsection{Data processing}
GI is a computational imaging technique, and it requires several computational steps to recover real-space images. Our implementation requires an additional step to decouple the structure of the incoming beam from the sample transmission, and to retrieve the normalization coefficients for the incoming beam intensity variations.
In this section, we describe this additional step, which is unique to our proposed setup. We refer to  Sec. C of the supp. mat. for the full GI data processing pipeline.
\\
If we assume that the sample transmission does not change throughout the GI acquisition, it is possible to decouple the sample transmission from the beam structure through principal component analysis (PCA).
Let us represent the sample transmission with the vector $\underline{f}$ (which also includes the spatial beam intensity profile), the beam structures for each GI realization as the set of vectors $M = \{ \underline{m}_j \}$, where $j \in [1, K]$ is the index of each GI realization and $K$ is the total number of realizations, and the beam intensity for each realization with the set of coefficients $C = \{ c_j \}$.
For each realization, the transmission of the sample plus the structured beam is $\underline{t}_j = c_j \underline{f} \otimes \underline{m}_j$, where $\otimes$ is the element-wise vector product. The average of all transmission images is $\frac{1}{K} \sum_j^K \{ c_j \underline{f} \otimes \underline{m}_j \} = \underline{f} \otimes \frac{1}{K} \sum_i^N c_j \underline{m}_j$.
% For absorption contrast, $0 \leq m_{ji} \leq 1$ where $i$ is the pixel in the recorded beam structure $j$.
The patterns in the set $M$ are supposed to be uncorrelated with each other, and $\underline{f}$ is the only component present in each transmission image $\underline{t}_j$. Thus, it is the dominant component in the PCA of the matrix $T = [ \underline{t}_1 , \underline{t}_2 , \dots , \underline{t}_K ]$. If we split the highest PCA component from the others, and reconstruct these two sets separately, we will obtain $c_j \underline{f}$ (called eigen-flat-field) from the former set, and $\underline{m}_j$ from the latter set, for each GI realization $j$.
\\
We show an example of transmission $\underline{t}$ in~\cref{fig:eigff:transmission}, the corresponding eigen-flat-field in~\cref{fig:eigff:flatfield} and beam structure in~\cref{fig:eigff:clean}.
The intensity fluctuations $c_j$ are computed by integrating each eigen-flat-field $c_j \underline{f}$, where $\sum_i f_i$ with $i \in [1, N]$ is the sum of all the pixels $i$ in the image $\underline{f}$, and resulting in a constant multiplicative factor. The integrated intensities in~\cref{fig:data:norm} are in good agreement with the low-frequency trend of the corresponding Fe $K_\alpha$ values in~\cref{fig:data:before}. We normalize the values in~\cref{fig:data:before} by the computed intensities in~\cref{fig:data:norm}, subtract the mean, and obtain the corrected GI realization intensities in~\cref{fig:data:after}.
\\
The GI reconstruction pixel size is determined through the auto-correlation (AC) function of the structured beams~\cite{Kingston2020}. Supposing that the mask is not periodic and sampled over non-overlapping regions, the resulting structured beams are not correlated with each other. Thus, it is enough to compute the AC of each structured beam with itself.
%This is done by first computing their 2D normalized AC function, and then performing azimuthal integration around the origin. The resulting 1D function is the isotropic AC function (of each structured beam).
The selected GI reconstruction pixel size is the lowest half-width half-maximum (HWHM) of all the AC curves. More details can be found in the supplemental material.
In the presented experiment, we found a minimum HWHM equal to 4 imaging detector pixels: Given its pixel size of $6 \, \upmu$m, it is equivalent to a $24 \, \upmu$m GI pixel size. Therefore, we binned the mask images $4 \times 4$, as shown in~\cref{fig:eigff:binned}.
\\
The GI reconstruction is performed with the Primal-Dual Hybrid Gradient (PDHG) algorithm, and using the $l_1$-norm minimization of the isotropic Total Variation (TV) of the image~\cite{Sidky2012}. The weight of the TV term is selected through cross-validation~\cite{Lane2020}.
The data processing code can be found at~\cite{Vigano2019b}.

\subsection{Results}
The XRF spectrum of the analyzed sample is shown in~\cref{fig:res:spectrum}. The Fe and Cu peaks from the wires are clearly visible. The C $K_\alpha$ line of the plastic capillary is below the detection limit of the XRF detector, but lines of other elements can be seen, e.g. Zn.
\begin{figure}[t]
    \centering
    \begin{subfigure}[b]{0.49\linewidth}
        \setlength{\abovecaptionskip}{4pt}
        \setlength{\belowcaptionskip}{-2pt}
        \includegraphics[width=\linewidth]{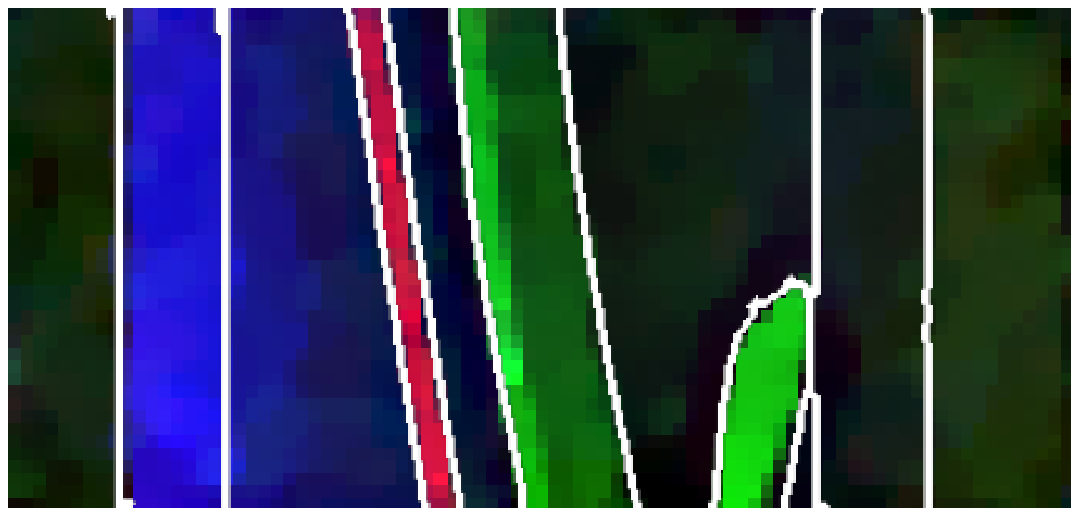}
        \subcaption{Fe-Cu-Zn / sample edges}
        \label{fig:recs:composite}
    \end{subfigure}
    \hfill
    \begin{subfigure}[b]{0.49\linewidth}
        \setlength{\abovecaptionskip}{4pt}
        \setlength{\belowcaptionskip}{-2pt}
        \includegraphics[width=\linewidth]{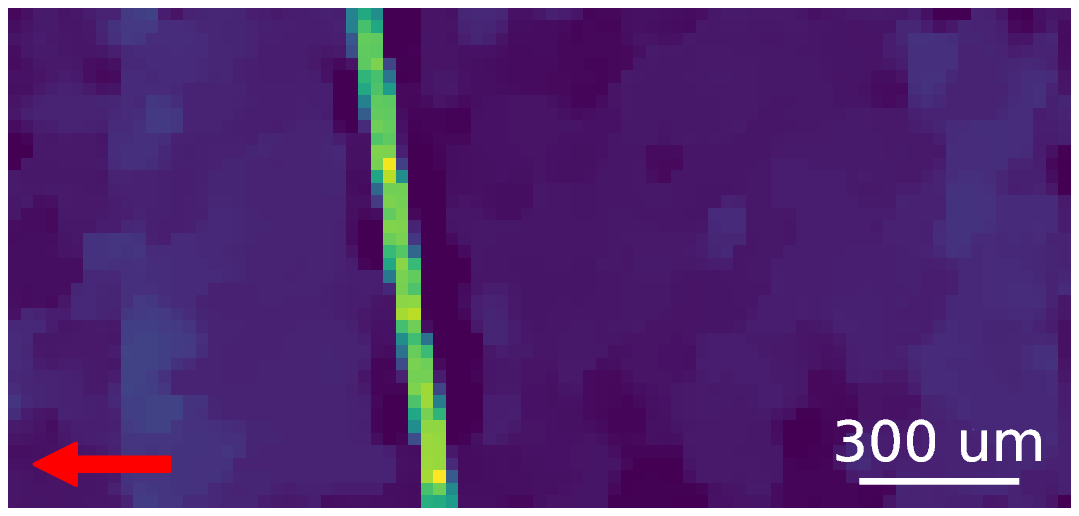}
        \subcaption{Fe}
        \label{fig:recs:Fe}
    \end{subfigure}
    \\
    \vspace{0.2cm}
    \begin{subfigure}[b]{0.49\linewidth}
        \setlength{\abovecaptionskip}{4pt}
        \setlength{\belowcaptionskip}{-2pt}
        \includegraphics[width=\linewidth]{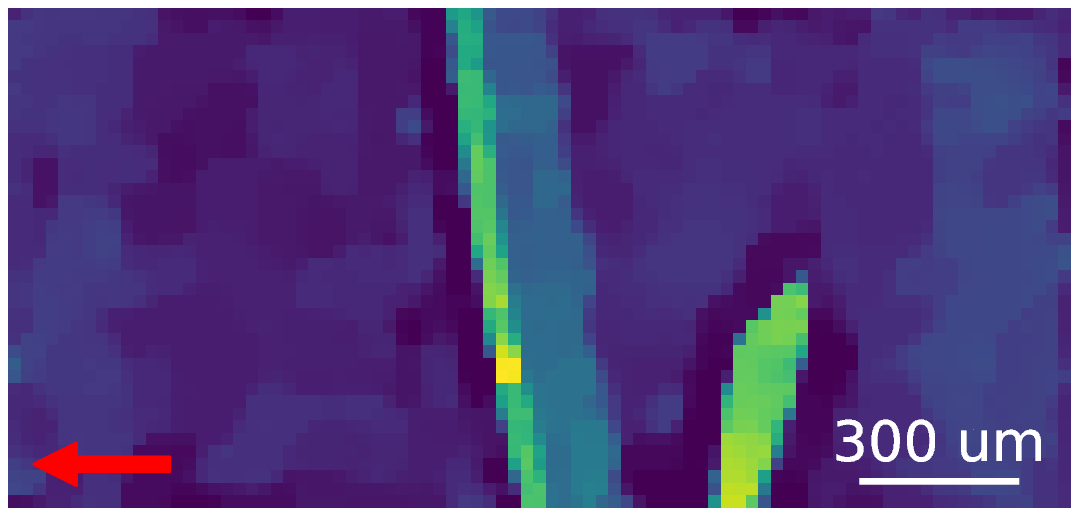}
        \subcaption{Cu}
        \label{fig:recs:Cu}
    \end{subfigure}
    \hfill
    \begin{subfigure}[b]{0.49\linewidth}
        \setlength{\abovecaptionskip}{4pt}
        \setlength{\belowcaptionskip}{-2pt}
        \includegraphics[width=\linewidth]{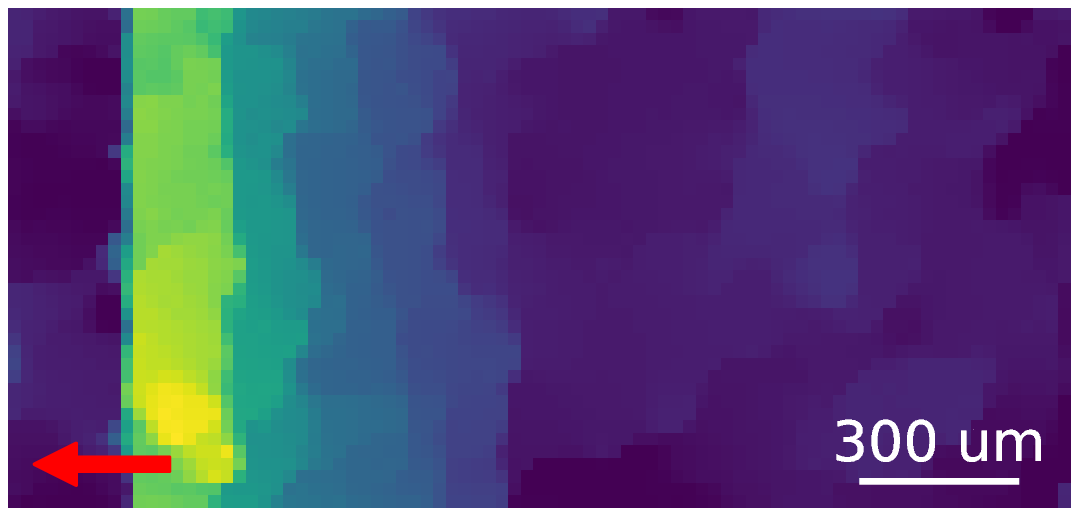}
        \subcaption{Zn}
        \label{fig:recs:Zn}
    \end{subfigure}

    \caption{(\textbf{\subref{fig:recs:composite}})~Composite image of Fe, Cu, Zn (Red, Green, and Blue respectively), and super-imposed edges of the transmission image in~\cref{fig:eigff:flatfield}; and the qualitative elemental reconstructions of: (\textbf{\subref{fig:recs:Fe}})~Fe; (\textbf{\subref{fig:recs:Cu}})~Cu; and, (\textbf{\subref{fig:recs:Zn}})~Zn. The images are affected by self-attenuation of the XRF photons. The XRF detector position is indicated with a red arrow.}
    \label{fig:recs}
\end{figure}
\\
% In \cref{fig:recs:composite}, we present the transmission image of the sample (directly acquired, not reconstructed), normalized by the open-beam intensity. It presents the position and shape of the objects in the FoV: The smaller Fe wire (left) and the two Cu flattened wires (center and right) are in center of the FoV, while the plastic capillary wraps all around them. This transmission image is morphologically identical to the retrieved eigen-flat-field (minus the open-beam intensity normalization), suggesting that the structured beams are correctly retrieved. 
% \\
In~\cref{fig:recs}, we find the GI reconstructions of the Fe, Cu, and Zn XRF signals in the FoV (\cref{fig:recs:Fe}, \cref{fig:recs:Cu}, and~\cref{fig:recs:Zn} respectively).
In~\cref{fig:recs:composite}, we present a color-coded composite image of the Fe, Cu, and Zn XRF signals (Red, Green, and Blue respectively), with super-imposed edges from~\cref{fig:eigff:flatfield}. This figure shows that the elemental images are in good agreement with the corresponding expected objects in the transmission image: The spatial distribution of the elements is correctly recovered (aside from artifacts).
The two main types of artifacts are: (a) trace signals from other channels; and (b) self-attenuation.
Point (a) means that the signal for one XRF line can have long tails in its energy dispersion function, and its signal can be mistakenly associated with other XRF lines. As a result, reconstructions show faint features from other elements. For instance, the Fe reconstruction (\cref{fig:recs:Fe}) shows traces of the capillary on the leftmost part of the FoV.
Point (b) means that the sample attenuates some of the XRF photons that it emits. As a result, regions of the sample further away from the detector have a much lower signal than regions of the sample closer to the XRF detector. This effect is clearly visible in both the Cu and Zn reconstructions (\cref{fig:recs:Cu}, and \cref{fig:recs:Zn}). More specifically, the biggest Cu wire is reconstructed with a higher signal on its left side, and only the leftmost part of the capillary is reconstructed. The XRF detector is in fact positioned on the left side of the images, as seen by the transmission detector (indicated with a red arrow in~\crefrange{fig:recs:Fe}{fig:recs:Zn}). This type of artifact is common to all XRF methods.
\\
The reconstructed images are approximately $1 \times 2 ~ mm^2$ in size, which corresponds to $42 \times 87$ pixels (for $24 \, \upmu$m pixel size). The acquisition consisted of $16 \times 56$ GI realizations, which represent $\sim 24.52 \%$ of the reconstructed pixels.

\subsection{Impact \& Outlook}
This proof-of-concept experiment overcomes important limitations associated with a synchrotron-based GI implementation. It demonstrates XRF-GI on a synchrotron beamline, that is not otherwise suited for XRF imaging.
This paves the way to enabling GI advantages (e.g. reduced dose) to synchrotron-based applications. It also enables the study of samples that could not be imaged with a PB approach, because of radiation damage or the inability to move them. This includes samples that would be perturbed by the back-and-forth translation of PB acquisitions (e.g. liquids), or with heavy sample environments that could not be displaced with enough precision or speed.
\\
In addition, the proposed XRF-GI implementation is robust against mask positioning errors or drifts, because the structure of the beam is acquired in line with the sample (without beam splitting), and simultaneously with the XRF signal. This is not the case in other GI acquisition setups. Thus, it implies a simplified setup, with fewer sources of noise or misalignment, compared to beam-splitter-based setups or computational GI approaches.
% Optionally, the masks can be mapped beforehand with a higher spatial resolution or longer exposure. The inline images obtained during the XRF-GI acquisition can then be used to register the high-quality masks with the XRF-GI data.
Compared to PB scanning, our XRF-GI implementation is also robust against image distortions arising from drifts during an entire acquisition. The sample is visible behind the mask at each realization, thus its position can be tracked and corrected. This could have its biggest impact at the nano-scale, where tracking drifts and positioning errors is of vital importance. % to reduce image distortions.
% But it could also be applied back to laboratory sources~\cite{Klein2022}, to increase their precision and reliability.
\\
Enabling synchrotron-based XRF-GI has a direct impact on every technique that leverages the XRF signal. As an example, XAS (x-ray absorption spectroscopy) often uses XRF as a high-quality proxy for the absorption signal. With XRF-GI, we enable obtaining spatial information from XAS measurements, without radically changing a beamline layout with respect to beam shaping and sample positioning.
% XAS measurements require fast energy scanning of the incoming beam, and could benefit from relatively larger incident beams, to increase the XRF yield. PB scanning would require introducing slits and focusing optics in the beam, and ensure mechanical precision much finer than the beam size.
% It would dramatically increase the acquisition time, by simultaneously introducing spatial scanning and reducing the illumination cross-section.
XAS measurements with PB scanning would experience dramatically increased acquisition time.
An XRF-GI-based implementation would instead enable probing of large regions of the sample at once, with much fewer realizations (i.e. scanning points) than the number of pixels (by leveraging compressive sensing~\cite{Lane2020}). % a much-reduced positioning precision than the PB counterpart, and with 
This would unlock routinely obtaining spatially resolved XAS maps within the time bounds imposed by beamtime allocation. Its applications span from the study of microscopic inclusion in macroscopic naturally occurring samples, to the observation of localized charge transfer phenomena in batteries.
% \\
% Our XRF-GI implementation only requires minor changes to pre-existing setups: Adding a two-dimensional detector downstream of the sample, and a translation stage for scanning the masks. Importantly, the translation of the masks does not require high precision or reproducibility.
%A PB setup would instead require introducing slits and focusing optics in the beam, and  ensure mechanical precision much finer than the beam size.
% In conclusion, XRF-GI will open new possibilities in existing beamlines and techniques, that were not originally designed to provide spatial XRF information. 
\\
%For synchrotron applications, XRF-GI can be specifically used to mitigate localized radiation damage to the samples. For dose sensitive samples, XRF-GI's reduced dose deposition per unit-surface per unit-time means that the heat and damage are more evenly distributed over the FoV. In turn, this can result in easier cooling and smaller deformation of the samples through-out the scans. This is especially important for the increased fluxes of 4th generation synchrotrons.
%\\
% XRF-GI can also be used in conjunction with XRF PB scanning. GI could deliver a preliminary low-resolution map of large regions of the sample, with much fewer exposures than the number of pixels, by leveraging compressive sensing (CS)~\cite{Lane2020}. This would allow to more quickly identify selected regions of interest to be then scanned with high-resolution and high signal-to-noise ratio PB scans.
% \\
% Finally, the presented XRF-GI setup paves the way to the implementation of GI in the synchrotron for all emission-contrast based imaging techniques (e.g. Compton, Auger, etc). Identical or very similar setups could be envisioned for those techniques, and their implementation would be compatible with existing spectroscopy setups.
% \\
In its current implementation, our technique requires that both the object and the mask are transparent enough to be both visible on the imaging detector, at the same time. This limitation can be relaxed, by scanning the mask beforehand. Even when some regions of the sample were to fully attenuate the transmitted beam, the transmission of one region of the mask in each GI realization would be enough to retrieve the correct position of the mask. This would also relax the dependency on the eigen-flat-field extraction step (i.e. using PCA), especially in the presence of sample drifts during the scan.

\begin{backmatter}
\bmsection{Funding} PAZY Foundation; Australian Research Council (Discovery project, DP210101312).

\bmsection{Acknowledgments} AK acknowledges support from the Australian Research Council through funding of the Discovery Project DP210101312. This research was supported by the Pazy Foundation. NV acknowledges Emmanuel Brun for lending the masks, and K. Joost Batenburg, Giovanni O. Lepore, and Daniele Pelliccia for the fruitful discussion.
%Additional information crediting individuals who contributed to the work being reported, clarifying who received funding from a particular source, or other information that does not fit the criteria for the funding block may also be included

% \bmsection{Disclosures} Disclosures should be listed in a separate section at the end of the manuscript. List the Disclosures codes identified on the \href{https://opg.optica.org/submit/review/conflicts-interest-policy.cfm}{Conflict of Interest policy page}. If there are no disclosures, then list ``The authors declare no conflicts of interest.''

% \smallskip

% \noindent Here are examples of disclosures:

\bmsection{Disclosures} The authors declare no conflicts of interest.

% \bmsection{Data Availability Statement} A Data Availability Statement (DAS) will be required for all submissions beginning 1 March 2021. The DAS should be an unnumbered separate section titled ``Data Availability'' that
% immediately follows the Disclosures section. See the \href{https://opg.optica.org/submit/review/data-availability-policy.cfm}{Data Availability Statement policy page} for more information.

% \begin{enumerate}
% \item When datasets are included as integral supplementary material in the paper, they must be declared (e.g., as "Dataset 1" following our current supplementary materials policy) and cited in the DAS, and should appear in the references.

\bmsection{Data availability} Data underlying the results are available at~\cite{Manni2023}. % presented in this paper 

% \item When datasets are cited but not submitted as integral supplementary material, they must be cited in the DAS and should appear in the references.

% \bmsection{Data availability} Data underlying the results presented in this paper are available in Ref. [3].

\bmsection{Supplemental document}
See Supplement 1 for supporting content.

\end{backmatter}

% \section{References}

% Note that \emph{Optics Letters} and \emph{Optica} short articles use an abbreviated reference style. Citations to journal articles should omit the article title and final page number; this abbreviated reference style is produced automatically when the \emph{Optics Letters} journal option is selected in the template, if you are using a .bib file for your references.

% However, full references (to aid the editor and reviewers) must be included as well on a fifth informational page that will not count against page length; again this will be produced automatically if you are using a .bib file.

% \bigskip
% \noindent Add citations manually or use BibTeX. See \cite{Zhang:14,OPTICA,FORSTER2007,testthesis,manga_rao_single_2007}.

% Bibliography
\bibliography{references}

\begin{thebibliography}{10}
\newcommand{\enquote}[1]{``#1''}

\bibitem{cit-material-science}
J.~A. van Bokhoven, T.-L. Lee, M.~Drakopoulos, C.~Lamberti, S.~Thie{\ss}, and
  J.~Zegenhagen, \enquote{{Determining the aluminium occupancy on the active
  T-sites in zeolites using X-ray standing waves},}
  {\protect\JournalTitle{Nature Materials}} \textbf{7}, 551--555 (2008).

\bibitem{cit-batteries}
F.~T. Haase, A.~Bergmann, T.~E. Jones, J.~Timoshenko, A.~Herzog, H.~S. Jeon,
  C.~Rettenmaier, and B.~R. Cuenya, \enquote{{Size effects and active state
  formation of cobalt oxide nanoparticles during the oxygen evolution
  reaction},} {\protect\JournalTitle{Nature Energy}} \textbf{7}, 765--773
  (2022).

\bibitem{cit-cultural-heritage}
R.~Ploeger and A.~Shugar, \enquote{{Where science meets art},}
  {\protect\JournalTitle{Science}} \textbf{354}, 826--828 (2016).

\bibitem{Sole2007}
V.~Sol{\'{e}}, E.~Papillon, M.~Cotte, P.~Walter, and J.~Susini, \enquote{{A
  multiplatform code for the analysis of energy-dispersive X-ray fluorescence
  spectra},} {\protect\JournalTitle{Spectrochimica Acta Part B: Atomic
  Spectroscopy}} \textbf{62}, 63--68 (2007).

\bibitem{daSilva2017}
J.~{Cesar da Silva}, A.~Pacureanu, Y.~Yang, S.~Bohic, C.~Morawe, R.~Barrett,
  and P.~Cloetens, \enquote{{Efficient concentration of high-energy x-rays for
  diffraction-limited imaging resolution},} {\protect\JournalTitle{Optica}}
  \textbf{4}, 492 (2017).

\bibitem{Vasin2007}
M.~Vasin, Y.~Ignatiev, A.~Lakhtikov, A.~Morovov, and V.~Nazarov,
  \enquote{{Energy-resolved X-ray imaging},}
  {\protect\JournalTitle{Spectrochimica Acta Part B: Atomic Spectroscopy}}
  \textbf{62}, 648--653 (2007).

\bibitem{Soltau2023}
J.~Soltau, P.~Meyer, R.~Hartmann, L.~Str{\"{u}}der, H.~Soltau, and T.~Salditt,
  \enquote{{Full-field x-ray fluorescence imaging using a Fresnel zone plate
  coded aperture},} {\protect\JournalTitle{Optica}} \textbf{10}, 127 (2023).

\bibitem{Moreau2018}
P.-A. Moreau, E.~Toninelli, T.~Gregory, and M.~J. Padgett, \enquote{Ghost
  imaging using optical correlations,} {\protect\JournalTitle{Laser \&
  Photonics Reviews}} \textbf{12}, 1700143 (2018).

\bibitem{Lane2020}
T.~J. Lane and D.~Ratner, \enquote{{What are the advantages of ghost imaging?
  Multiplexing for x-ray and electron imaging},} {\protect\JournalTitle{Optics
  Express}} \textbf{28}, 5898 (2020).

\bibitem{Klein2022}
Y.~Klein, O.~Sefi, H.~Schwartz, and S.~Shwartz, \enquote{{Chemical element
  mapping by x-ray computational ghost fluorescence},}
  {\protect\JournalTitle{Optica}} \textbf{9}, 63 (2022).

\bibitem{Bennink2002}
R.~S. Bennink, S.~J. Bentley, and R.~W. Boyd, \enquote{“two-photon”
  coincidence imaging with a classical source,} {\protect\JournalTitle{Physical
  Review Letters}} \textbf{89}, 113601 (2002).

\bibitem{Pelliccia2016}
D.~Pelliccia, A.~Rack, M.~Scheel, V.~Cantelli, and D.~M. Paganin,
  \enquote{{Experimental X-Ray Ghost Imaging},} {\protect\JournalTitle{Physical
  Review Letters}} \textbf{117}, 113902 (2016).

\bibitem{Gatti2004}
A.~Gatti, E.~Brambilla, M.~Bache, and L.~A. Lugiato, \enquote{Ghost imaging
  with thermal light: Comparing entanglement and classical correlation,}
  {\protect\JournalTitle{Physical Review Letters}} \textbf{93}, 093602 (2004).

\bibitem{Kingston2020}
A.~M. Kingston, G.~R. Myers, D.~Pelliccia, F.~Salvemini, J.~J. Bevitt,
  U.~Garbe, and D.~M. Paganin, \enquote{{Neutron ghost imaging},}
  {\protect\JournalTitle{Physical Review A}} \textbf{101}, 053844 (2020).

\bibitem{Sidky2012}
E.~Y. Sidky, J.~H. J{\o}rgensen, and X.~Pan, \enquote{Convex optimization
  problem prototyping for image reconstruction in computed tomography with the
  {C}hambolle-{P}ock algorithm,} {\protect\JournalTitle{Physics in Medicine and
  Biology}} \textbf{57}, 3065--3091 (2012).

\bibitem{Vigano2019b}
\url{https://github.com/cicwi/PyCorrectedEmissionCT}.

\bibitem{Manni2023}
\url{https://doi.org/10.5281/zenodo.7828494}.

\end{thebibliography}

% Full bibliography added automatically for Optics Letters submissions; the following line will simply be ignored if submitting to other journals.
% Note that this extra page will not count against page length
\bibliographyfullrefs{references}

\end{document}